\documentclass{aa}  

\usepackage{graphicx}
\usepackage{ulem}
\usepackage{latexsym}
\usepackage{natbib}
\bibpunct{(}{)}{;}{a}{}{,} 
\usepackage{txfonts}
\newcommand{\msun}{\mbox{$M_{\odot}$}}
\renewcommand{\d}{{\rm d}} 

\begin{document}
   \title{ACS Imaging of star clusters in M51\thanks{Based on observations made with the NASA/ESA \textit{Hubble Space Telescope}, obtained from the data archive at the Space Telescope Science Institute. STScI is operated by the association of Universities for Research in Astronomy, Inc., under the NASA contract NAS 5-26555.}}

   \subtitle{II. The luminosity function and mass function across the disk}

   \author{M.R. Haas
          \inst{1,2}
          \and
          M. Gieles \inst{3}
	  \and
	  R.A. Scheepmaker \inst{2}
	  \and
	  S.S. Larsen \inst{2}
	  \and
	  H.J.G.L.M. Lamers \inst{2}
          }

   \offprints{M.R. Haas, \email{haas@strw.leidenuniv.nl}}

   \institute{	Leiden Observatory, Leiden University, 	P.O. Box 9513, NL-2300 Leiden, The Netherlands
        \and	 
		Astronomical Institute, Utrecht University, Princetonplein 5, NL-3584 CC Utrecht, The Netherlands
	\and
		European Southern Observatory, Casilla 19001, Santiago 19, Chile
             }

   \date{Received ; accepted }

  \abstract
   {Whether or not there exists a physical upper mass limit for star clusters is as yet unclear. For small cluster samples the mass function may not be sampled all the way to the truncation, if there is one. Data for the rich cluster population in the interacting galaxy M51 enables us to investigate this in more detail.}
   {Using \textit{HST/ACS} data, we investigate whether the cluster luminosity function (LF) in M51 shows evidence for an upper limit to the mass function. The variations of the cluster luminosity function parameters with position on the disk are addressed.}
   {We determine the cluster LF for all clusters in M51 falling within our selection criteria, as well as for several subsets of the sample. In that way we can determine the properties of the cluster population as a function of galactocentric distance and background intensity. By comparing observed and simulated LFs we can constrain the underlying cluster initial mass function and/or cluster disruption parameters. A physical upper mass limit for star clusters will appear as a bend dividing two power law parts in the LF, if the cluster sample is large enough to sample the full range of cluster masses. The location of the bend in the LF is indicative of the value of the upper mass limit. The slopes of the power laws are an interplay between upper mass limits, disruption times and evolutionary fading.}
   {The LF of the cluster population of M51 is better described by a double power law than by a single power law. We show that the cluster initial mass function is likely to be truncated at the high mass end. We conclude from the variation of the LF parameters with galactocentric distance that both the upper mass limit and the cluster disruption parameters are likely to be a function of position in the galactic disk. At higher galactocentric distances the maximum mass is lower, cluster disruption slower, or both.}
   {}


   \maketitle
%

\section{Introduction}

The distinction between star clusters and galaxies used to be fairly clear, but recently the gap between both has been slowly filled up with objects in the intermediate mass range. The most massive star clusters were the Milky Way's old globular clusters, with masses going up to a few $10^6 \,\msun$. The least massive galaxies known were galaxies of the type of Fornax, having masses of about $10^7 \,\msun$. Many objects are now determined to be in between these distinct mass ranges, e.g. `Ultra Compact Dwarfs' \citep[e.g.][]{mieske04, hilker99}, `stellar superclusters' \citep[e.g.][]{hilker99} and `Dwarf-Globular Transition Objects' \citep[e.g.][]{hasegan05}. Whether these are all different names for the same objects or not remains to be seen.

Are these new classes of objects really formed in distinct formation processes, or is there rather a continuous transition between all, and are star clusters just a more common mode of formation? \citet{kissler-patig06} put young massive star clusters ($M = 10^{6.5 - 7}\,\msun$) on the scaling relations (relations between mass, velocity dispersion and radius) of other hot stellar systems and find that these clusters do follow scaling relations derived for the least massive galaxies, i.e. dwarf elliptical galaxies. Less massive clusters do not show a mass-radius relation. Simulations by e.g. \citet{fellhauerkroupa05} and \citet{bekki04} support the idea that the most massive cluster-like objects may have formed from merging smaller clusters, thereby establishing a mass-radius relation \citep{kissler-patig06}.

As this would imply that the formation scenarios of systems considerably more massive than $\sim 10^6 \,\msun$ are really distinct from `normal' star clusters, it is interesting to see whether we can also find evidence for a maximum mass for the star cluster-like objects. As pointed out by e.g. \citet{jordan07} and \citet[and references therein]{waters06}, the universality of the globular cluster luminosity function (GCLF) of globular clusters in other galaxies is much easier to achieve when including a fundamental upper mass limit, in the form of an exponential cut-off, for star clusters in dynamical cluster evolution models. This also holds for the globular cluster system of the Milky Way \citep[e.g.][]{burkertsmith00}.  The universal turn-over  
in the globular cluster mass function is achieved more easily with  
cluster evolution models when including a exponential cut off in the  
mass function around $10^6$ \msun.

To obtain detailed information (for example the mass, age and extinction) of a single cluster of which the stars are not resolved requires photometry in a wide range of broadband filters (UV to IR, e.g. \cite{degrijs05}, and references therein), or spectroscopy. In the study of star cluster populations, where spectroscopy is a very time consuming process, the cluster LF is a useful tool. This is especially true when photometry is only available in a small number of passbands and obtaining the mass, metallicity and extinction of separate clusters is hampered by the age-metallicity-extinction degeneracy. In such cases, the LF can be used as a statistical tool, giving only information about the population as a whole. 

In various environments the LF can be approximated well by a power law distribution ($N \, \d L \propto L^{-\alpha}\, \d L$), with a slope ($-\alpha$) between -1.8 and -2.4 (e.g. \citet{larsen02,degrijs03a}). The translation from the LF to the cluster initial mass function  
(cluster IMF) is non-trivial, however, especially when clusters spanning a range of ages are present. During its lifetime a cluster fades due to stellar evolution \citep[e.g.][]{schulz02} and loses stars due to a variety of disruption effects (e.g. \citet{baumgardtmakino03,lamers05,gieles06c,gieles06b}). Nevertheless, using assumptions for the cluster formation rate, the cluster IMF and cluster disruption parameters, one can construct expected LFs and subsequently compare these with observed LFs \citep{gieles05a}.

In three galaxies the LF of the rich star cluster systems appears to be better described by a double power law, i.e. two distinct parts, both described by a power law, which are separated by a bend at some absolute magnitude. This bend magnitude differs from galaxy to galaxy. \citet{whitmore99} found for the ``Antennae'' (NGC 4038/4039) a bend at $M_V \simeq -10$ (standard UBVI), with on the faint side a shallower slope ($\sim -1.7$) than on the bright side ($\sim -2.7$). For M51, \citet{gieles06a} found hints for a double power law behavior of the LF for a sub-sample of clusters in the inner part of the disk, which were later confirmed by \citet{gieles06} for a 5 times larger sample, based on the same data as used in this study. The slopes on both sides are similar to the slopes found by \citet{whitmore99}, but the bend occurs about 1.6 magnitudes fainter. With the distance modulus of the Antennae newly derived by \citet{saviane08}, which differs by 1.7 magnitudes from the older determination by \citet{whitmoreschweizer95}, the difference in bend magnitudes between M51 and the Antennae reduces to 0.1 magnitude. In \citet{gieles06a} it is shown that the LF of the star cluster population of NGC 6946 is also better fit with a double power law, with parameters comparable to M51. Note that the slopes at the faint end of the LF for all these galaxies are similar to the slopes found for populations with a single power law distribution. If the observed or formed number of clusters is not high enough to sample the LF up to the point where the bend occurs, the bend will not be detected. The galaxies for which a double power law LF for their clusters has been reported are all typified by a high star formation rate.

Whereas the bend in the LF of the ``Antennae'' was originally interpreted as being caused by a bend in the mass function, \citet{zhangfall99} already mentioned a truncated mass function as a possible explanation for the double power law behavior. \citet{gieles06a} have shown with analytic cluster population models that such a bend can occur if the highest occurring cluster mass is no longer determined by the size of the sample (i.e. the highest mass present in a cluster population is determined statistically).  For example, in the case of the LMC and SMC \citep{hunter03}, the cluster IMF does not show signs of an upper mass limit, and therefore sampling this mass function results in a maximum mass in the sample that is determined by the number of clusters and the slope of the power law cluster IMF. This is argued to be generally true by \citet{weidner04}, who argue that the maximum cluster mass is a function of the star formation rate only. On the other hand, if there exists a \textit{physical upper mass limit} for star clusters and the cluster IMF is sampled up to this limit the LF will appear in a shape more closely resembling a double power law \citep{larsen06}. 

If the cluster IMF is truncated, the physical reason for the maximum mass may have its imprints on the LF of subsets of the population. Variations of the maximum mass with galactocentric distance or surrounding ambient densities can hold important clues about the formation and evolution of star clusters.

Using deep observations by \textit{HST} using the \textit{Advanced Camera for Surveys (ACS)} of the interacting, face-on, spiral galaxy M51, which cover a large fraction of the disk of the galaxy and have a resolution of 0$\farcs$05, we are able to select a sample of thousands of clusters \citep{scheepmaker06}. Those clusters range in luminosity from the magnitude of single bright stars (distinguished on the basis of their spatial extent) to the brightest clusters found in spiral galaxies. The size of this sample gives us the opportunity to divide the clusters in smaller subsets and still have enough statistics for the obtained LFs. In this way we are able to study trends with e.g. galactocentric distance and the spiral arm structure of the galaxy. The same observations are used by \citet{hwanglee08} to determine the LF of luminous clusters. They find a single power law, but only include clusters brighter than the bend magnitude, as found by \citet{gieles06a} and \citet{gieles06}, with a power law slope which agrees with the bright end slope of \citet{gieles06a, gieles06}.

In \S~\ref{sec:obs} we describe the data, source selection and photometry.  The luminosity function of clusters in M51 will be presented in \S~\ref{sec:lumfunc}, for the whole population as well as for certain subsets. In \S~\ref{sec:modcomp} we perform Monte Carlo simulations of star cluster populations in order to investigate the dependence of obtained LF parameters on maximum cluster mass, disruption time and cluster formation history. There, we also draw qualitative conclusions about the cluster IMF and cluster disruption parameters of M51. The variation of the bend luminosity with position on the disk is further investigated in \S~\ref{sec:discussion} and our conclusions are summarized in \S~\ref{sec:conclusions}.


\section{Observations, photometry and sample selection} \label{sec:obs}

\subsection{Data}
We used the \textit{HST/ACS} Hubble Heritage data of M51 (NGC 5194) and its companion NGC 5195 in \textit{F435W}, \textit{F555W}, \textit{F814W ($\sim$B, V} and \textit{I}, respectively) and \textit{F658N (H$\alpha$)}, see Table~\ref{tab:data} (taken from \citet{mutchler05}, proposal ID 10452, PI: S.V.W. Beckwith). The six pointings correspond to a total field of view of 430'' x 610''. This corresponds to 17.5 x 24.8 kpc at a distance of 8.4 Mpc \citep{feldmeier97}, covering a large fraction of the disk (more than 2 scale lengths of $\sim 4$ kpc, \citet{beckman96}) of M51 plus the region with NGC 5195. This field is considerably larger than the fields of previous studies \citep{bastian05, gieles05a}.

For details on reduction we refer to \citet{mutchler05}. The same data have also been used for our previous study of the luminosity function of star clusters in M51 \citep{gieles06} and in a study on the radii of these clusters \citep{scheepmaker06}.

For a detailed description of the data analysis we refer to \citet{scheepmaker06}. In view of the importance of the sample selection for this study a concise summary is given below.

\begin{table}
\caption{The data set of the observations of M51, taken under proposal ID 10452.}             
\label{tab:data}      
\centering                          
\begin{tabular}{c c c}        
\hline\hline                 
Filter & Exposure time (s) & Limiting magnitude  \\    
\hline                        
F435W (B) & 4 x 680 = 2720 & 27.3 m$_\textrm{B}$ \\      
F555W (V) & 4 x 340 = 1360 & 26.5 m$_\textrm{V}$ \\
F814W (I) & 4 x 340 = 1360 & 25.8 m$_\textrm{I}$ \\
F658N (H$\alpha$, [N II]) & 4 x 680 = 2720 & - \\
\hline                                   
\end{tabular}
\end{table}


\subsection{Source selection, photometry and radius measurements}
Selection of pointlike sources was done with the SExtractor package \citep{bertinarnouts96}. 
Photometry was performed using the \textit{IRAF/DAOphot} package. All magnitudes will be quoted in the Vega magnitude system. An aperture correction was applied, with the aid of artificial sources. The aperture corrections were all done for 3 pc sources. Since size and luminosity are generally found to be only weakly correlated for star clusters \citep{zepf99, larsen04, bastian05, scheepmaker06}, we do not expect the use of a constant aperture correction to introduce significant biases in our LF estimation. All clusters are affected by the same Galactic foreground extinction. We used the tables of \citet{schlegel98}. For accurate local (i.e. in the M51 system) extinction determinations one needs a wide range of broadband photometry, in order to overcome the age/metallicity/extinction degeneracy. Because we only have \textit{B,V,I} photometry we are not able to clearly distinguish between the three effects. Therefore we \textit{can not} correct for local extinction. 

We used the \textit{ISHAPE} routine within the \textit{BAOlab} package \citep{larsen99,larsen04a} to determine the effective radii (projected half light radii) of all point-like sources. Analytic cluster profiles were convolved with an empirical PSF of the camera and fit to the sources. For high S/N ($> 30$) sources, sizes measured by ISHAPE on HST images are typically accurate to $\sim$0.2 pixels, so we adopt this as the lower limit for an object we can recognize as resolved  \citep{larsen04a}. With \textit{ACS}, at the distance of M51, this corresponds to 0.5 pc. We therefore take this as a lower limit. For a detailed investigation of the radii of this cluster population, see \citet{scheepmaker06}.

\subsection{Background regions} \label{sec:background}
\begin{figure}
   \centering
   \includegraphics[width=85mm]{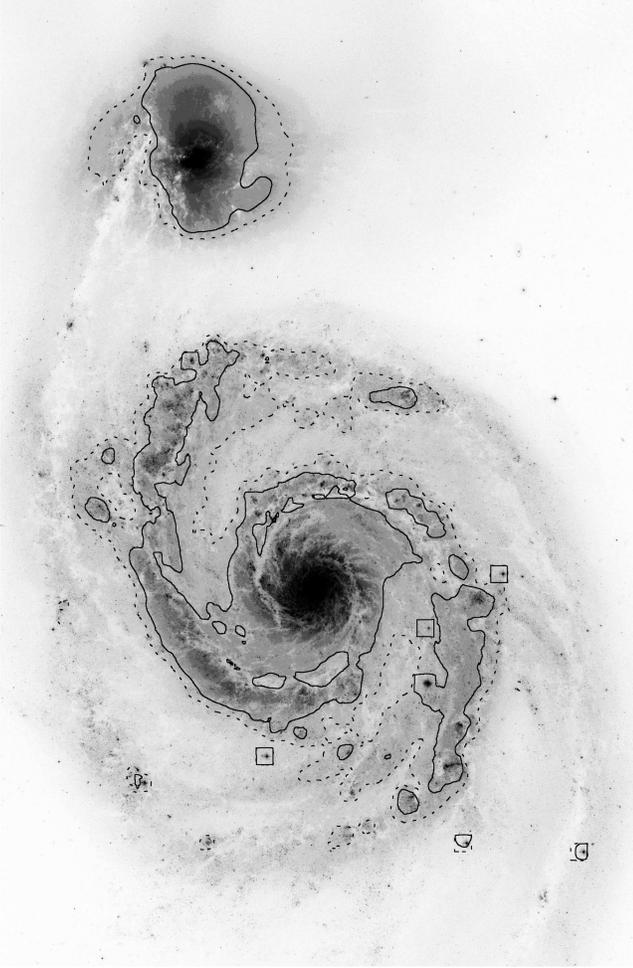}
      \caption{The contours that outline the three background intensity level regions, superimposed on the image in the \textit{F555W} passband. The solid line encloses the highest background level and everything outside the dotted line is called low background. The area in between the two lines is referred to as being a transition region, to have the other two regions clearly distinguishable.}
         \label{fig:mask}
\end{figure}

Because the background intensity varies strongly over the image, we divided the image in three background levels, as indicated in Fig.~\ref{fig:mask}. The image has been smoothed with a boxcar average of 200 pixels. Two isophotes on this smoothed image are used as background limits (corresponding to 20.72 and 21.02 mag arcsec$^{-2}$, respectively). We will use these three regions when investigating trends in the LF with background intensity. The high background region will also be referred to as being the spiral arms, whereas the low background will be called interarm region.

\subsection{Completeness} \label{sec:completeness}
Completeness tests were performed with the aid of artificial cluster experiments. In this study we make use of the completeness limits for 3 pc sources (see Table~\ref{tab:completeness}), determined in a high background region, to be sure that our sample is not affected by incompleteness due to high surface brightness. As a slope of a luminosity function might be sensitive to an incompleteness of $\sim 10$\%, it is worth noting that using a higher completeness limit cuts off a big fraction of the clusters. As can be found in \citet{scheepmaker06}, the completeness function becomes very flat above 90\%. An estimate of the expected bias in determining the slope of the low luminosity end of the LF in the three filters can be found in Section~\ref{sec:lumfunc}.

\begin{table}
\caption{90 and 95\% completeness limits for the three broad passbands. The same limits are used for different distance and background regions. They are all determined for a 3 pc cluster in a high background intensity region.}             
\label{tab:completeness}      
\centering                          
\begin{tabular}{l c c}        
\hline\hline                 
Filter & 90\% completeness & 95\% completeness\\    
       
\hline                        
\textit{F435W} ($\sim B$)& 24.2 & 23.6 \\
\textit{F555W} ($\sim V$)& 23.8 & 23.0 \\
\textit{F814W} ($\sim I$)& 22.7 & 22.2 \\
\hline                                   
\end{tabular}
\end{table}

\subsection{Sample selection}
Finally, sources were selected from the complete sample (i.e. the sample of clusters, brighter than the 90\% completeness limit) if:
\begin{enumerate}
\item the source is detected in \textit{F435W}, \textit{F555W} and \textit{F814W},
\item the source was fit better with an extended cluster profile than with a delta function, meaning that the $\chi_{\nu}^2$ of the extended profile is lower than that of a pure PSF fit, in the filter of interest (this results in slightly different cluster samples in different filters),
\item the cluster has a radius (in the case of an elliptical cluster profile, the semi-major axis is converted to the radius of a circle with the same area) of at least 0.5 pc,
\item the nearest neighboring point source is at least 5 pixels away to reduce the contamination.
\end{enumerate} 

\noindent The resulting sample contains 7698, 6846 and 2539 clusters for \textit{F435W, F555W} and \textit{F814W}, respectively. To be sure we use a reliable sample of clusters we visually inspected the sample brighter than the determined completeness limits that pass the criteria above. For a description of the criteria to remove clusters from the sample and a graphical representation of clusters that are removed by visual inspection, see \citet{scheepmaker06}.
Note that the sample used in this study still contains more sources than the sample in the study described by \citet{scheepmaker06}, since there they need to be sure that they have a reliable estimate for the radius of the source, whereas in this study the only constraint is that we need to be sure that the source is extended.


\section{The luminosity function of clusters in M51} \label{sec:lumfunc}
Following the source selection as described in Section~\ref{sec:obs}, we obtain the luminosity function of the star cluster candidates in M51. The distributions are fit with single and double power laws. In Appendix~\ref{sec:fits} we compare several methods for determining the shape of the LF. Following the results we will use a method that is described and tested by \citet{maizapellaniz05}, based on \citet{dagostinostephens}. In short, it uses bins that are variable in width, such that every bin contains an equal number of clusters. This avoids differences in statistical weights when fitting the distribution function, which would bias the obtained slope of the power law to bins with high statistical weight. The power law LF will be defined as
\begin{equation}
\frac{\d N}{\d L_{\lambda}} \propto L_{\lambda}^{-\alpha}
\end{equation}
or, equivalently
\begin{equation} \label{eq:dplf}
\frac{\d N}{\d M_{\lambda}} \propto 10^{\beta \cdot M_{\lambda}}
\end{equation}
where $L_{\lambda} \, (M_{\lambda})$ is the luminosity (magnitude), $N$ is the number of clusters and $\alpha$ and $\beta$ are related as $\alpha~=~2.5~\cdot~\beta~+~1$.

Equation \ref{eq:dplf} is applied in fits for single power laws. For double power laws we apply
\begin{equation}
\frac{\d N}{\d M_{\lambda}} = \left \{ 
\begin{array}{lcr}
A \cdot 10^{M_{\lambda} \cdot \beta_1} & \, {\rm for} & M_{\lambda} > M_{\rm bend} \\
A \cdot 10^{((\beta_1 - \beta_2) \cdot M_{\rm bend})} \cdot 10^{M_{\lambda} \cdot \beta_2} & \, {\rm for} & M_{\lambda} \leq M_{\rm bend} \\
\end{array}
\right .
\end{equation}
in which both power law slopes, the location of the bend ($M_{\rm bend}$) and the normalization are free parameters. Note that only the normalization of one power law is a free parameter, the other one is then set by the location of the bend and that normalization. The total number of free parameters for the single power law is 2, while for the double power law fits there are 4 free parameters. The number of bins, and therefore the number of degrees of freedom, is different for all samples. The number of bins ($N_{\rm bins}$) is related to the number of objects in the sample ($N_c$) by $N_{\rm bins} = 2 \cdot N_c^{2/5} + 50$ \citep[adapted from][]{dagostinostephens}.

The fit is performed on the whole range from the 90\% completeness limit to the brightest cluster in the sample. As we use the 90\% completeness limits, we will here estimate by how much we might underestimate the slope of the low luminosity end due to incompleteness. By correcting the LF at the 90 and 95\% limits for the incompleteness the LF would become steeper. If we were to correct the value of the LF at the 90 and 95\% completeness limits, the change in $\beta$ would be $\Delta \beta = 2.5 / \Delta M_\lambda \cdot (\log(1/0.9) - \log(1/0.95)$, in which $\Delta M_\lambda$ is the difference between the magnitudes of the 90 and 95\% completeness limits per filter. As $\Delta \alpha = 2.5 \cdot \Delta \beta$, we obtain the following estimates for the underestimate of the low luminosity end slope of the LF, for \textit{F435W}, \textit{F555W} and \textit{F814W} respectively: 0.098, 0.073 and 0.12. Note that these estimates are upper limits, as the 95\% completeness limits always falls well within the faint end of the LF, as will become clear in the following section. The underestimate only holds for the range in magnitudes between the 90 and 95\% completeness limits, brightward of the 95\% limit the underestimate is less severe.

\subsection{Luminosity functions of the whole sample}
\begin{figure} 
   \centering
   \includegraphics[width=\columnwidth]{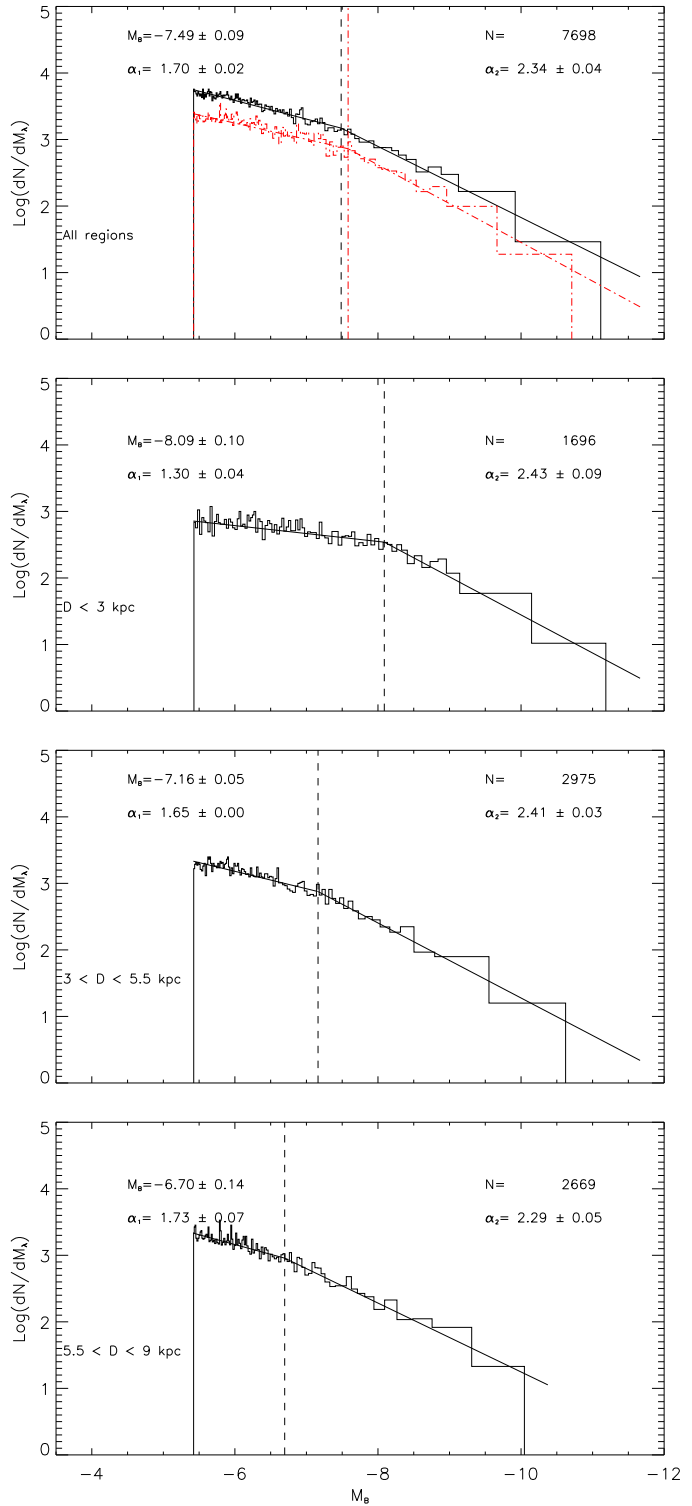}
      \vspace{0.3cm} \caption{The LF (number of clusters divided by the bin width) of the identified clusters in M51 in $M_{F435W}$. The double power law fits are performed on all of these clusters. Only clusters with a galactocentric distance $<$ 9 kpc are in this sample. The number of clusters in the sample is indicated in all panels. The top panel is the whole sample. The lower three are samples, in number more or less equally divided, at different galactocentric radii. Both slopes and the magnitude of the bend are indicated (vertical dashed line). The dot-dashed line in the upper panel is the LF for clusters with $r_{eff} > 2 pc$.}
         \label{fig:lfbd}
\end{figure}
   
\begin{figure}
   \centering
   \includegraphics[width=\columnwidth]{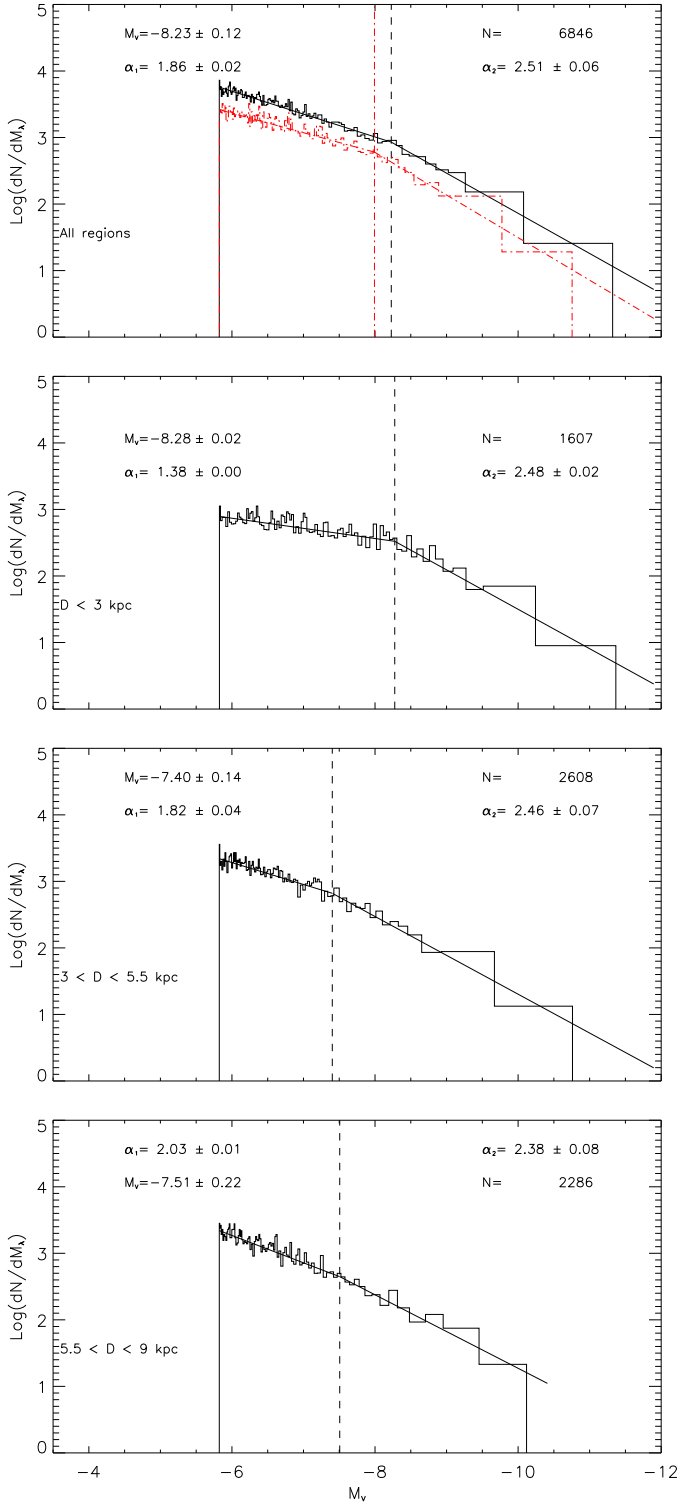}
      \vspace{0.3cm} \caption{Same as Fig.~\ref{fig:lfbd}, but now for \textit{F555W}.}
         \label{fig:lfvd}
\end{figure}

\begin{figure}
   \centering
   \includegraphics[width=\columnwidth]{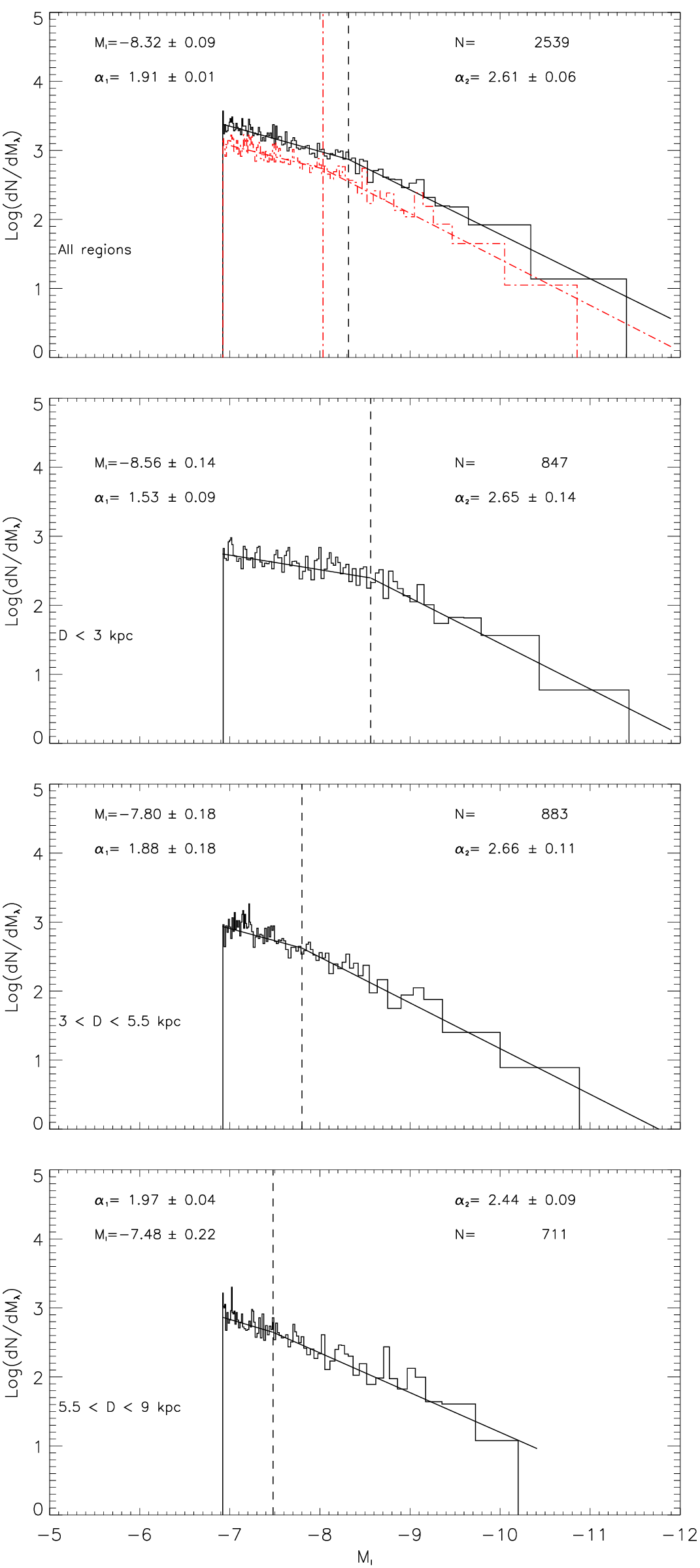}
      \vspace{0.3cm} \caption{Same as Fig.~\ref{fig:lfbd}, but now for \textit{F814W}.}
         \label{fig:lfid}
\end{figure}

For \textit{F435W}, \textit{F555W} and \textit{F814W} the LFs are shown in the top panels of Fig.~\ref{fig:lfbd}, \ref{fig:lfvd} and \ref{fig:lfid} respectively. They are fit with a double power law distribution function. This fits the data better than a single power law, as already found by \citet{gieles06}. 

To confirm their conclusion, the fits are also performed with a single power law distribution function, using the fit method as described in the Appendix. In order to compare the fits, a comparison of the (reduced) $\chi^2_\nu$ is made. There are two more parameters: an extra slope and a bend magnitude. Note however, that these parameters are not independent, as the slope of one of the power laws, together with the location of the bend, define at least one point in the other power law (they meet at the bend location). The difference in fit quality is indicated by both reduced $\chi^2$ ($\chi^2_\nu$) in the final two columns, together with the number of degrees of freedom.

In the upper panels of Fig.~\ref{fig:lfbd}, \ref{fig:lfvd} and \ref{fig:lfid} we also show the LF for all clusters with an effective radius bigger than 2 pc using dot-dashed lines. The LF parameters agree within 1$\sigma$ with the LF for all clusters with $r_{eff} > 0.5$ pc. This reassures that the sample is not contaminated with bright stars.

For an overview of the fit results, see Table~\ref{tab:fits}. Faintward of the bend magnitude the slope is shallower than brightward of the bend as expected from the artificial cluster populations presented in \citet{gieles06a} and in agreement with previous observations \citep{whitmore99,gieles06}.

\citet{hwanglee08} used the same data set, but different cluster identification criteria, to determine the LF of the same population of clusters. Comparing their single power law fit and the double power law fit as found by \citet{gieles06}, they conclude that a single power law gives a better fit, by comparing $\chi^2$ for both fits on the \textit{F555W} image. \citet{hwanglee08} found for \textit{F555W} on the interval $-10 < M_{F555W} < -8$, a single power law  with power law index $-2.59 \pm 0.03$. This is  consistent with our high luminosity end slope of $-2.52 \pm 0.06$, which is found for $M_{F555W} < -8.23$. The bend luminosity found in \citet{gieles06} is 0.7 mag brighter in \textit{F555W} than the bend found in this work. The reasons are a correction for average extinction (0.3 mag in \textit{F555W}), a different binning method and slightly different cluster selection criteria in \citet{gieles06}.

\begin{table*}
\caption{Fit results of the whole sample in all three passbands. Every cluster brighter than the completeness limit in the filter under consideration is taken into account in the fits. Single power law as well as double power laws are fit, to judge the statistical validity of the bend. The first column is the passband, the second the number of clusters within the fit range. Column three contains the slope of the single power law fit, whereas the fourth, fifth, sixth and seventh column contain both the slopes and the location of the bend of the double power law respectively. The final two columns show the reduced $\chi^2$ for single ($\chi^2_{\nu,s}$) and double ($\chi^2_{\nu,d}$) power law fits, respectively, with the number of degrees of freedom indicated between brackets. All errors are $1 \sigma$.}
\label{tab:fits}      
\centering          
\begin{tabular}{l l l l l l l l l}     
\hline\hline       
  &   & \multicolumn{1}{l}{Single power law} &  \multicolumn{4}{l}{Double power law} \\
Passband & N  &  $\alpha$ & $\alpha_1$ & $\alpha_2$ & $M_{\textrm{bend}}$ & $\chi^2_{\nu,s}$ (\#d.o.f.) & $\chi^2_{\nu,d}$ (\#d.o.f.) \\
\hline                    
$F435W$ & 7698 & 1.98 $\pm$ 0.01 & 1.70 $\pm$ 0.02 & 2.34 $\pm$ 0.04 & -7.49 $\pm$ 0.09 & 3.16 (119) & 0.96 (117) \\
$F555W$ & 6846 & 2.07 $\pm$ 0.01 & 1.86 $\pm$ 0.02 & 2.51 $\pm$ 0.06 & -8.23 $\pm$ 0.12 & 2.41 (116) & 0.99 (114) \\
$F814W$ & 2539 & 2.28 $\pm$ 0.02 & 1.92 $\pm$ 0.01 & 2.61 $\pm$ 0.06 & -8.32 $\pm$ 0.09 & 1.50 (94) & 0.88 (92) \\
\hline                  
\end{tabular}
\end{table*}

\begin{table*}
\caption{Fit results of the subsets in galactocentric distance in all three passbands. For every distance bin, the best fit faint- and bright end slope ($\alpha_1$ and $\alpha_2$, respectively) as well as the magnitude of the bend are indicated. All errors are $1 \sigma$. The reduced $\chi^2$ values for both single and double power law fits (and the number of degrees of freedom) are given in the last column.}
\label{tab:lffarvsclose}      
\centering          
\begin{tabular}{l l l l l l l l} 
\hline\hline       

 Passband & $D_G$ &  N & $\alpha_1$ & $\alpha_2$ & $M_{\textrm{bend}}$ & $\chi^2_{\nu,s}$ (\#d.o.f.) & $\chi^2_{\nu,d}$ (\#d.o.f.)  \\

\hline                     
$F435W$& $D_G < 3$ kpc & 1696 & 1.30 $\pm$ 0.04 & 2.43 $\pm$ 0.09 & -8.09 $\pm$ 0.10  & 3.34 (87) & 1.00 (85) \\
 & $ 3 < D_G < 5.5$ kpc  & 2975 & 1.65 $\pm$ 0.00 & 2.41 $\pm$ 0.03 & -7.16 $\pm$ 0.05  & 2.11 (97) & 0.87 (95) \\
 & $ 5.5 < D_G < 9$ kpc & 2669 & 1.73 $\pm$ 0.07 & 2.29 $\pm$ 0.05 & -6.70 $\pm$ 0.14   & 1.39 (94) & 0.94 (92) \\
\hline
$F555W$ & $D_G < 3$ kpc & 1607 & 1.38 $\pm$ 0.00 & 2.48 $\pm$ 0.02 & -8.28 $\pm$ 0.02   & 2.87 (86) & 0.96 (84) \\ 
 & $ 3 < D_G < 5.5$ kpc & 2608 & 1.82 $\pm$ 0.04 & 2.46 $\pm$ 0.07 & -7.40 $\pm$ 0.14   & 1.75 (94) & 1.09 (92) \\
 & $ 5.5 < D_G < 9$ kpc & 2286 & 2.03 $\pm$ 0.01 & 2.38 $\pm$ 0.08 & -7.51 $\pm$ 0.22   & 1.19 (92) & 1.08 (90) \\
\hline
$F814W$ & $D_G < 3$ kpc & 847 & 1.53 $\pm$ 0.09 & 2.65 $\pm$ 0.14 & -8.56 $\pm$ 0.14   & 1.57 (77) & 0.84 (75) \\
 & $ 3 < D_G < 5.5$ kpc & 883 & 1.88 $\pm$ 0.18 & 2.66 $\pm$ 0.11 & -7.80 $\pm$ 0.18   & 1.12 (78) & 0.93 (76) \\
 & $ 5.5 < D_G < 9$ kpc & 711 & 1.97 $\pm$ 0.04 & 2.44 $\pm$ 0.09 & -7.48 $\pm$ 0.22   & 0.81 (75) & 0.82 (73) \\
\hline                  
\end{tabular}
\end{table*}

\begin{table*}
\caption{Fit results of the subsets in the high and low background regions in all three passbands. For both backgrounds, the best fit faint- and bright end slope ($\alpha_1$ and $\alpha_2$, respectively) as well as the magnitude of the bend are indicated. The final two columns show the reduced $\chi^2$ for single ($\chi^2_{\nu,s}$) and double ($\chi^2_{\nu,d}$) power law fits, respectively, with the number of degrees of freedom indicated between brackets. All errors are $1 \sigma$.}             
\label{tab:lfhighvslow}      
\centering          
\begin{tabular}{l l l l l l l l} 
\hline\hline       

 Passband & Background & N & $\alpha_1$ & $\alpha_2$ & $M_{\textrm{bend}}$ & $\chi^2_{\nu,s}$ (\#d.o.f.) & $\chi^2_{\nu,d}$ (\#d.o.f.) \\

\hline                     
$F435W$& Low & 2837 & 1.99 $\pm$ 0.06 & 2.49 $\pm$ 0.07 & -6.91 $\pm$ 0.17   & 5.28 (96) & 1.02 (94) \\
       & High& 2771 & 1.10 $\pm$ 0.04 & 2.24 $\pm$ 0.05 & -7.41 $\pm$ 0.07   & 1.19 (95) & 0.83 (93) \\
\hline
$F555W$ & Low & 2378 & 2.19 $\pm$ 0.08 & 2.54 $\pm$ 0.08 & -7.43 $\pm$ 0.29   & 4.62 (92) & 1.00 (90) \\ 
       & High & 2606 & 1.38 $\pm$ 0.07 & 2.41 $\pm$ 0.07 & -8.00 $\pm$ 0.08   & 1.21 (94) & 1.11 (92) \\
\hline
$F814W$ & Low & 623 & 1.98 $\pm$ 0.33 & 2.61 $\pm$ 0.10 & -7.30 $\pm$ 0.22   & 1.57 (74) & 0.86 (72) \\
       & High & 1333 & 1.60 $\pm$ 0.09 & 2.50 $\pm$ 0.08 & -8.20 $\pm$ 0.13   & 0.75 (83) & 0.73 (81) \\

\hline                  
\end{tabular}
\end{table*}

\subsection{Variations with galactocentric distance}
The total sample is large enough to cut it in three, more or less equally sized (in number of clusters), galactocentric distance `bins', without losing too much statistical confidence. The limits are chosen at 3 and 5.5 kpc. For the distance dependencies of LF parameters, only clusters within a circle with galactocentric distance smaller than 9 kpc are used, in order to be sure that all the clusters belong to M51 and not to its companion NGC5195. In the resulting three regions the luminosity functions are shown in Fig.~\ref{fig:lfbd}, Fig.~\ref{fig:lfvd} and Fig.~\ref{fig:lfid} for \textit{F435W}, \textit{F555W} and \textit{F814W} respectively. The results of the double power law fits are shown in Table~\ref{tab:lffarvsclose}.

We find that the location of the bend shifts to fainter magnitudes for greater galactocentric distance for all filters. The bend magnitudes in the different filters get brighter for redder filters. 

Furthermore, in all three filters, the slope of the faint end of the LF gets steeper for subsets further out in the galaxy. In the $D < 3$ kpc subsamples it is even clearer that a double power law fits the distribution better, as the difference between slopes is biggest here.

\begin{figure}
   \centering
   \includegraphics[width=\columnwidth]{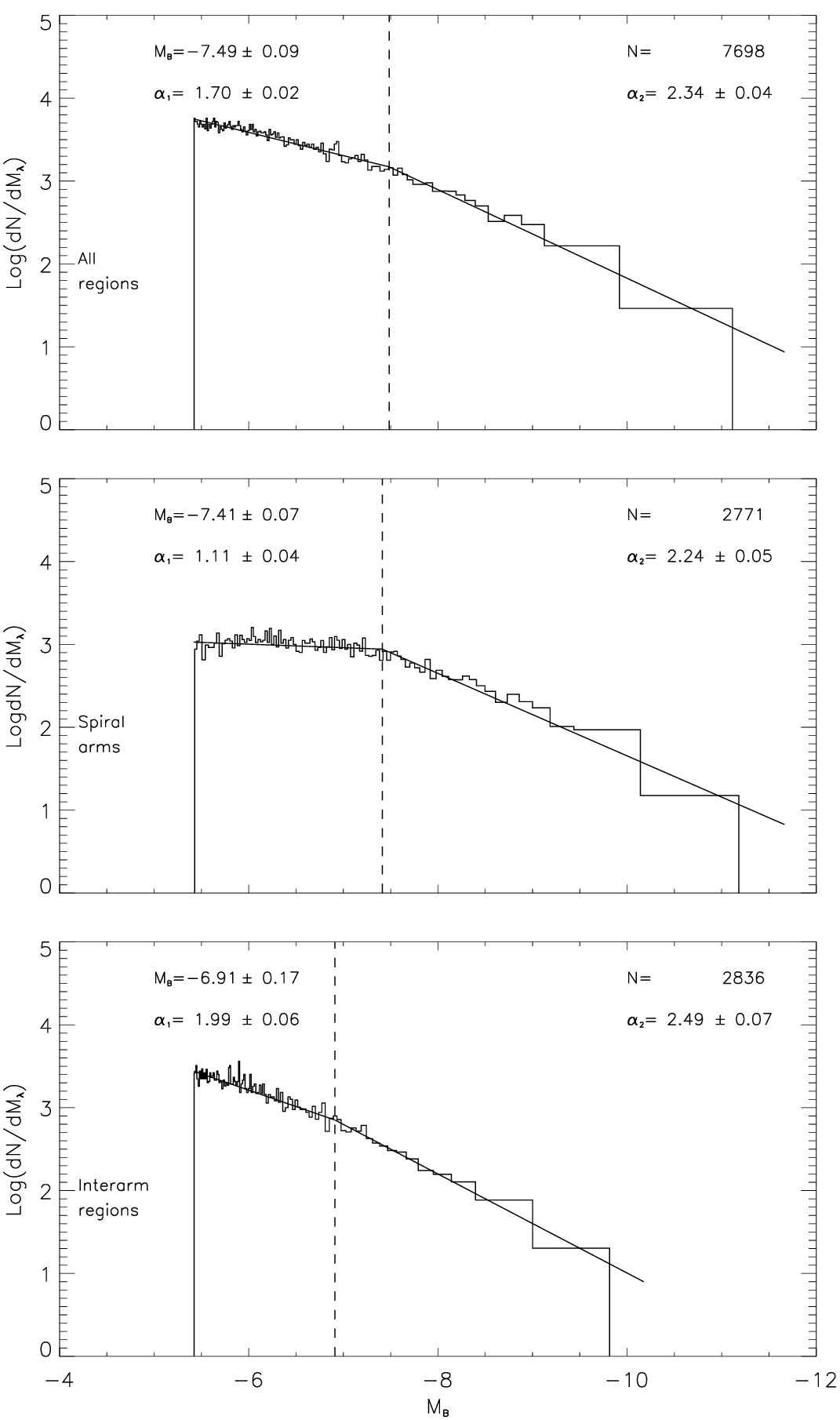}
      \caption{Same as Fig.~\ref{fig:lfbd}, but now with varying background intensity, comparing spiral arms with interarm regions.}
         \label{fig:lfbr}
\end{figure}

\subsection{Variations with background intensity}
The background intensity regions, as described in Sect.~\ref{sec:background}, also divide the sample in sub-samples, in which we can look for variations in the shape of the LF. The results are shown in Fig.~\ref{fig:lfbr} for the \textit{F435W} passband. Because the results are very similar for different filters, we do not show the LF for other filters, but we do list all results in Table~\ref{tab:lfhighvslow}. Only the high and low background regions are compared.  

The location of the bend is about half a magnitude brighter in higher background regions, although uncertainties are bigger than for the galactocentric distance dependence. The faint-end slope depends quite strongly on background intensity. It is much shallower in high background regions, again showing clearly the difference in fit quality between single and double power law descriptions (which of course manifests itself more strongly in subsamples for which the faint- and bright-end slopes are strongly different). 

If clusters are formed in the spiral arms, a younger population of clusters would be expected in the high background intensity region. This is, for this sample of clusters, confirmed by \citet{scheepmaker06}. We would therefore expect to find in the high background region a faint-end slope of the LF, more closely resembling the assumed cluster IMF slope of -2. This slope is found in earlier studies \citep{bik03, bastian05, gieles06a}. This is \textit{not} what we observe. A possible explanation is the blending of faint sources. If one source is very close to another, and those are together selected as one source, then the light is added. Thereby, two fainter cluster disappear from the sample and one brighter cluster appears, which flattens the LF. It is difficult to estimate the detailed effect of blending on the LF, since the youngest clusters tend to be found within complexes \citep{bastian05a} and blending will be a strong function of cluster age, luminosity and environment. Although we have carried out completeness experiments in regions of high background, these might still be optimistic for the most extreme cases of crowding.

The trends with background intensity are correlated with the trends with galactocentric distance, because the high background regions are on average closer to the center than the regions of low background. It is hard to disentangle the two trends and judge which is the most fundamental of the two. We conclude that we find a significant dependence of the faint-end slope on whether the population of clusters is in a high or low background region and that the bend in the LF occurs roughly half a magnitude brighter in the high background intensity region.


\section{Shaping the LF of the M51 cluster population} \label{sec:modcomp}
To get a handle on the variation of the LF parameters across the disk, we will discuss the variation of the LF parameters of artificial cluster models. These models compute the integrated properties of a cluster population, such as the LF, age distribution and mass distribution, based on an assumed cluster formation history, initial cluster mass function and cluster disruption scenario, which are all variable. Fitting the observed LF parameters in the disk of M51 in detail requires fitting a large parameter space in cluster IMF, cluster formation history and disruption parameters as a function of time and position in the disk, especially for interacting galaxies, where star formation histories and local environmental conditions are generally complex. 

Here, we will investigate a range in cluster formation histories, disruption times and maximum masses in a Monte Carlo simulation.

\subsection{Artificial cluster populations}
If a certain cluster formation history and cluster IMF are assumed, one obtains a full history of cluster formation, which provides one with an age and initial mass for every cluster in the system. Together with a description for the cluster disruption (mass as a function of age for a cluster with certain initial mass), this gives at all moments the present day mass and age of a cluster. For given metallicity and stellar initial mass function, also the luminosity in every photometric passband is available. These come from stellar population synthesis models. The models of \citet{gieles05a}, that are used here, use the \textit{GALEV} models \citep{schulz02,andersfritzevalvensleben03}.

\subsubsection{Cluster formation histories} \label{sec:cfh}
We construct artificial populations for different cluster formation histories. The underlying smooth cluster formation history is taken to be a power law in time:
\begin{equation} \label{eq:cfh}
\frac{\d N}{\d t} \propto t^{p}
\end{equation}
where the parameter $p$ is taken zero (constant cluster formation in time), positive ($p=0.5$, for a cluster formation rate increasing with time, i.e. higher now than in the past) or negative ($p=-0.5$, for a declining cluster formation with time). The ages are drawn between a minimum and maximum age of $10^7$ and $10^{10}$ yr, respectively. On top of all these smooth formation histories we add an extra burst of cluster formation, forming an extra factor of 0 (no burst) or 0.1 (forming 10\% of the mass of the smooth population in the burst on top of the smooth formation). This value of 0.1 is chosen to be an extreme case, as  M51 and its companion are not really merging, but rather a system in a close encounter. The age of the clusters formed in the burst is taken to be $3 \cdot 10^{7}$ yr, the age at which \citet{bastian05} found a peak in the cluster formation history of this galaxy.

We assume all cluster populations to form with a cluster initial mass function that is described by a power law with index $-2$, truncated at the maximum mass (see below).

\subsubsection{Maximum masses and disruption times}
Disruption times are taken $\log(t_4 / \textrm{yr}) = [\infty, 9, 8]$ to investigate the difference between no disruption, moderate disruption and strong disruption. $t_4$ is the total disruption time of a cluster with an initial mass of $M_i = 10^4 \msun$. For other masses the disruption time is given by
      \begin{equation} \label{eq:tdis}
       t_{\rm dis} (M_i) = t_4 (M_i/10^4 \msun)^{\gamma}
      \end{equation}
with $\gamma=0.62$ \citep{boutloukoslamers03,baumgardtmakino03}. For the mass of a cluster as a function of age we take the gradual dissolution prescription of \citet{lamers05}, without taking into account the mass lost due to stellar evolution (because the population synthesis models use initial stellar mass as input). The maximum masses are varied between $10^{4.25}$ \msun\ and $10^{6}$ \msun\, taking steps of $\log(M) = 0.25$. The lower limit is chosen such that the clusters are still within the observable range, the upper limit goes up to the point where the mass function is not any longer sampled up to the maximum mass. This last point is disruption time dependent as will become clear in the following sections.

\subsubsection{Analysis of the model population}
For all cluster masses a maximum lifetime is determined from their initial mass and their corresponding disruption time (Eq.~\ref{eq:tdis}). For all surviving clusters we derive the magnitude, by scaling the magnitude from the \textit{GALEV} stellar population models with a \citet{kroupa01} stellar IMF for solar metallicity to the current mass. Because we do not have any information about internal extinction, we do not apply any attenuation on the cluster magnitudes. This may shift the bend magnitude of the LF to brighter magnitudes, by a few tenths of magnitudes, compared to the observations. Differential extinction may change the slopes of the power laws as well.

Because the resulting LF parameters from the fits will depend on the number of clusters observed (which determines how far the LF is sampled at the high luminosity end) we follow a very big population of clusters. We make sure that the population consists of more clusters than we have observed. When fitting the LF in the same way as the data, we only take clusters brighter than our 90\% completeness limits into account. In order to have the same statistics we now take a random sample of clusters from this population, exactly equal to the number of clusters observed in the specific filter under consideration. We repeat every run 20 times with a different random number seed.

We apply the exact same analyses as for the observed data, fitting both single and double power laws to the distribution of cluster luminosities. We keep track of the difference in fit quality and will not include populations that are fit better with a single power law (determined from $\chi_{\nu}^2$) or that have a better fit with a double power law, but with the bend at the edges of the range of luminosities in the sample (essentially also a single power law). We take the median bend magnitude of all the populations that satisfy these criteria. Furthermore, we only include realizations which sample the mass function up to the upper mass limit. If not sampled that far, one does not expect a bend in the LF and the bend that might be found does not mean much. Note that just a few runs are thrown out (5 out of 20 at most, 0 in 90\% of the $t_4-M_{\rm max}$ combinations).

\begin{figure}
   \centering
   \includegraphics[width=\columnwidth]{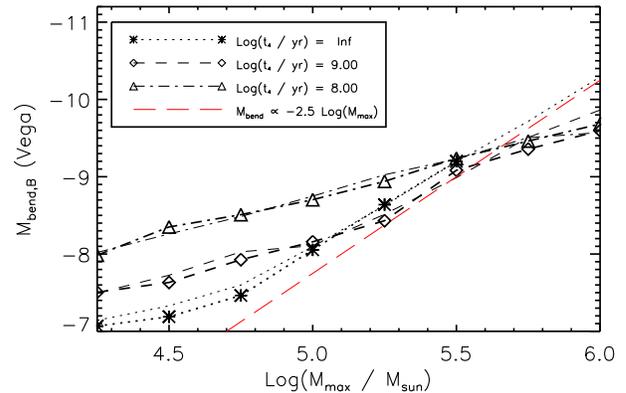}
      \caption{The bend magnitude as a function of maximum mass, for disruption times $\log(t_4 / \textrm{yr}) = [\infty, 9, 8]$. The cluster formation rate is taken constant in time and there is no extra burst included. The three different line styles correspond to different disruption times as stated in the legend. The thick line is a Monte Carlo realization with the number of clusters in sample equal to the observed number of clusters in the same filter. The thin lines (without symbols) are for the same disruption time, now with the number of clusters in the sample scaled with the maximum mass. The thin long-dashed line is the line for which $M_{\rm bend} \propto -2.5 \log(M_{\rm max})$, the relation expected on simple theoretical grounds, as explained in the text.}
         \label{fig:cstsfh}
\end{figure}

\subsection{Resulting LF parameters}
We will only show results for the \textit{F435W} filter, as here the statistics of the observations are best due to the longer integration time. Results are not qualitatively different in other filters, although the exact value of the slopes of the LF and the bend magnitude are different.

In Figure~\ref{fig:cstsfh} the resulting bend magnitude as a function of maximum cluster mass is shown for the three different disruption times, for a constant cluster formation rate, without any extra burst. It can be clearly seen that, in general, the bend shifts to brighter magnitudes for higher maximum masses. At high masses, the relation is cut-off, because the mass function is no longer sampled up to the maximum mass, when using the same number of clusters as observed. When repeating the experiment where we scale the number of clusters in the sample linearly with the maximum mass (thin lines in the figure), the upward trend is continued. In this case the mass function is always sampled to well above the upper mass limit.

\subsubsection{Explaining the dependence of the bend magnitude on maximum mass}
For a cluster sample with a uniform age distribution and no  disruption, the bend luminosity is that of a cluster of the maximum allowed mass, with an age equal to the age of the cluster sample \citep{gieles05a}. If disruption effects and non-uniform age distributions are allowed the relation between bend luminosity and maximum mass becomes more complicated, as discussed below.

\begin{figure}
   \centering
   \includegraphics[width=\columnwidth]{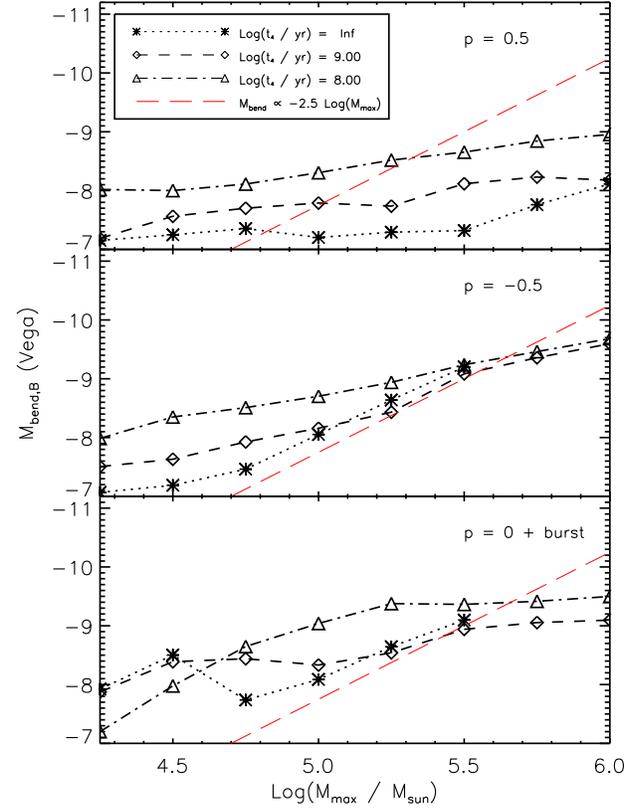}
      \caption{Same as Figure~\ref{fig:cstsfh}, but now for three different cluster formation histories. \textit{upper panel:} an increasing cluster formation rate with time ($\d N/\d t \propto t^{0.5}$) without burst, \textit{middle panel:} a decreasing cluster formation rate ($\d N/\d t \propto t^{-0.5}$), without burst and \textit{lower panel:} a constant cluster formation rate, as in Figure~\ref{fig:cstsfh} but with an extra burst, forming 10\% of the stars with an age of $3\cdot 10^7$ yr.}
         \label{fig:diffsfh}
\end{figure}

For no disruption, we can see in Figure~\ref{fig:cstsfh} (dotted line) that, if the maximum mass is considerably higher than the lower mass limit of visible clusters (depending on age between $10^3$ and $10^{\sim 4.5}$), the relation between maximum mass and bend magnitude is given by $M_{\rm bend} \propto -2.5 \log(M_{\max})$, as expected from simple arguments: luminosity depends linearly on mass for fixed age and metallicity, and the magnitude $M_{\lambda} \propto -2.5 \log(L_{\lambda})$. This relation is overplotted on Figs.~\ref{fig:cstsfh} and \ref{fig:diffsfh} with the long dashed lines.

\subsubsection{The effect of cluster disruption on LF parameters}
The main influence of cluster disruption is seen in the slope of the low-luminosity end of the LF. This is because the cluster disruption time depends on its mass, with low mass clusters disrupting faster than more massive clusters. Disrupting a cluster means moving it to a lower mass bin, so mass dependent disruption means that clusters are moved to a bin belonging to lower mass at an ever decreasing rate for increasing mass. \textit{Cluster disruption therefore flattens the LF and stronger disruption will lead to stronger flattening}. 

The bend location is also affected by the disruption time. The maximum mass occurring in a log(age) bin, now is not the maximum initial mass, but the maximum initial mass evolved in time due to disruption. In case of an untruncated mass function, this would just make the relation between the number of objects in a log(age) bin grow slower than linear with age, resulting in a shallower LF slope, than the slope of the cluster initial mass function. When adding a maximum mass, the highest possible mass is also a function of age. Stronger disruption lowers the mass of a cluster more during the lifetime of the galaxy. Therefore, for the same maximum mass the bend becomes fainter at the high maximum mass end of the Figure (\ref{fig:cstsfh}). For low maximum masses, a cluster of the maximum mass can be completely disrupted in the lifetime of a galaxy. In that case, the bend would occur at the luminosity of the most massive \textit{surviving} cluster at a relatively young age. In that case, the bend is brighter than it would be without disruption, see Fig.~\ref{fig:cstsfh}.
\textit{Summarizing this section we conclude that disruption makes the relation between the maximum mass and bend magnitude flatter than it would be without disruption.}

\subsubsection{The influence of the cluster formation history}
The factor that is probably worst constrained by the data is the cluster formation history. Here we will show that a difference in cluster formation, even in simple cases, already makes the obtained bend magnitude as a function of maximum mass very degenerate with disruption and maximum mass. We simulate a population of clusters with the power law cluster formation rate as a function of time given by Equation~\ref{eq:cfh} with $p=0.5$ and $p=-0.5$, to investigate the resulting bend magnitudes for cluster formation rates increasing and decreasing with time, respectively. We also investigate the effect of a starburst on top of a constant cluster formation history, as described in Section~\ref{sec:cfh}. The resulting maximum mass-bend magnitude relations are shown in Figure~\ref{fig:diffsfh}.

The behavior of the bend magnitude with maximum mass for a decreasing cluster formation rate with time (i.e. lower now than in the past) is very similar to that of a constant star formation history (Fig.~\ref{fig:cstsfh}), as is shown in the middle panel of Fig.~\ref{fig:diffsfh}. As the bend occurs at the magnitude of the highest mass cluster at the age of the population, this is expected, since in both cases clusters with relatively old ages receive strong weight in the combined LF of the whole sample. Therefore the bend magnitude, which is the magnitude below which clusters of all ages contribute to the LF, remains unchanged.

For a cluster formation rate which is increasing with time, the oldest clusters receive relatively little weight in the combined LF, and therefore the bend does not show up clearly at the expected magnitude. Instead, the LF is dominated by the combination of cluster disruption and the cluster formation history. For stronger disruption, the LF becomes more weighted towards younger clusters, so for stronger disruption the obtained bend is brighter.

An extra burst, as depicted in the lower panel of Figure~\ref{fig:diffsfh}, results in very irregular relations between bend magnitude and maximum mass. It depends strongly on the fraction of stars formed in the burst and the age of the burst. Therefore, without a very detailed knowledge about the burst, constraining the mass function or disruption is nearly impossible.

\subsubsection{The wavelength dependence of the bend}
The bend luminosity is expected to be higher in redder filters, as the shape of a clusters spectral energy distribution is dominated by red stars for clusters with the age of the population (i.e. old clusters). In red filters there is least evolution in the luminosity over the lifetime of a cluster. \textit{This results in a steeper bright-end slope for the red filter than for the blue filter}. 

As described in \citet{gieles08}, the bright end slope of the LF can be predicted and depends only on the fading of a cluster in a specific filter. If the luminosity of a cluster in a filter as a function of time can be approximated by a power law, as $L(\lambda, t) \propto t^{-\zeta}$, then the bright end LF slope should be $\d N / \d L_{\lambda, up} \propto L_{\lambda}^{-1-1/\zeta}$. Taking values for $\zeta$ from \citet{gieles07} (for clusters younger than 3 Gyr), we obtain the following predicted high luminosity end slopes for \textit{F435W}, \textit{F555W} and \textit{F814W}: -2.28, -2.52 and -2.85, respectively.

\subsection{The cluster population of M51}
On the basis of a comparison between, on the one hand, the qualitative explanation of LF parameters as a function of cluster IMF and cluster disruption given above, and on the other hand the observed trends listed in Tables \ref{tab:fits}, \ref{tab:lffarvsclose} and \ref{tab:lfhighvslow} the following qualitative interpretation of the observed LF parameters can be given:
\begin{enumerate}
\item \textit{The disruption time for star clusters varies with galactocentric distance, such that disruption is strongest in the central regions of the galaxy}. This conclusion can be drawn from the shallower faint-end side of the LF in the central regions of the galaxy. Concluding on the dependence of disruption time on spiral structure is hampered by the contradiction between expected and observed dependence of the faint-end slope of the LF on spiral structure. Besides, there is a degeneracy between variations as a function of galactocentric distance and background intensity that arises from the fact that high background intensity regions preferentially lie closer to the center of the galaxy than high background intensity regions. Selection effects, though, may be important.
\item Translating trends in bend magnitude with radius and background level
  to a trend in terms of maximum cluster mass is non-trivial and
  requires detailed knowledge about the formation and disruption  
histories
  of the cluster populations involved. We tentatively suggest that
  the observed trend of decreasing bend magnitude with galactocentric
  distance (and lower background level) may be indicative of a  
corresponding
  decrease in maximum cluster mass. Assuming a constant upper mass limit and simple cluster formation history would require large variations in the disruption time across the disk.
\end{enumerate} 

The conclusion on differences in disruption time between spiral arms and interarm regions might not hold if the age distribution in spiral arms is different from the age distribution in interarm regions. This is found to be the case by \citet{scheepmaker06}. The variation of the bend luminosity between arm and interarm regions might be explained by extinction, which is probably stronger in the spiral arm than in the interarm regions. With this limited set of broad band filters we can not test this hypothesis.


\section{The variation of the bend luminosity across the disk} \label{sec:discussion}
Here, we investigate the variation of the luminosity of the bend over the disk. At fixed age, the luminosity of a cluster scales linearly with its mass, and therefore the variation of the luminosity of the bend should follow the variation in maximum mass across the disk.

\begin{figure}
   \centering
   \includegraphics[width=\columnwidth]{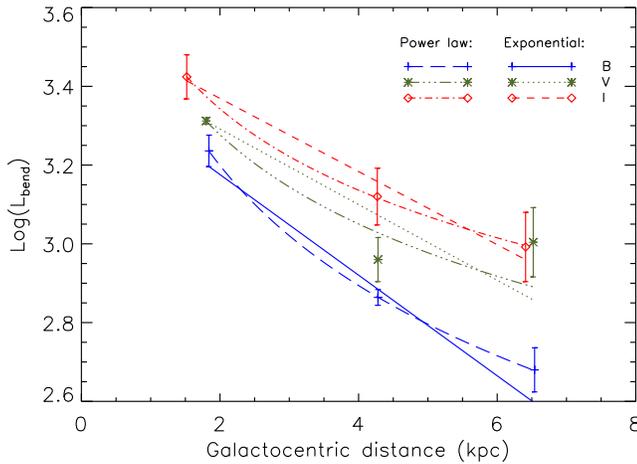}
      \caption{For all three filters, $-0.4\cdot M_{bend} = \log(L_{\rm bend}/L_0)$, with $L_0$ the filter dependent zero-point luminosity is plotted against the galactocentric distance. Exponential fits are shown for all three filters, as well as power law fits. The error bars are 1$\sigma$ errors.}
         \label{fig:varexp}
\end{figure}

As luminosity scales linearly with mass and magnitudes logarithmically, we will fit the quantity $L = 10^{-0.4 \cdot M_{bend}}$ as a function of galactocentric distance.
Fitting an exponential to the three points in galactocentric distance (the mean galactocentric distance of all clusters in the sample; note that this is different for the three filters, as the samples are not the same), versus $\log{L_{\rm bend}}$ (the logarithm of luminosity in units of the filter dependent zero point luminosity), gives scale lengths of 3.4$\pm 0.4$, 4.5$\pm 0.7$ and 4.7$\pm 1.0$ kpc for \textit{F435W, F555W} and \textit{F814W}, respectively. See Fig.~\ref{fig:varexp}. In these fits, the errors on the bend magnitudes are taken into account. The error on mean galactocentric radii for all sub-samples are assumed to be negligible (the error can be estimated to be a few pixels for every cluster, which results in relative errors much smaller than the relative errors on the bend magnitude). When fitting power laws to the points, the indices ($p$) of the power laws ($\log(L_{\rm bend}) \propto p \cdot \log(D_{gal})$) for the fits through \textit{F435W}, \textit{F555W} and \textit{F814W} are -1.0$\pm 0.1$, -0.8$\pm 0.1$ and -0.7$\pm 0.1$ respectively. These fits are also shown in Fig.~\ref{fig:varexp}

Because we have only three points per filter, and because of the lack of detailed modelling of the cluster population (and therefore the lack of precise determinations of the maximum mass), we do not want to discriminate between a power law dependence and an exponential decline of the bend luminosity with galactocentric distance.


\section{Conclusions} \label{sec:conclusions}
Using high resolution, deep \textit{HST/ACS} imaging, we have confirmed the fact that the luminosity function of star clusters in M51 is better described by a double power law distribution function than by a single power law. Although any sub-sample of clusters is fit only slightly better by a double power law than by a single power law, the fact that it holds for \textit{all} sub-samples strengthens the belief that the LF is indeed fitted better with a double power law than with a single power law. The fits on small galactocentric distance bins and high background regions clearly show that a single power law is insufficient in describing the LF. Also, the fact that for every subsample the luminosity of the break is higher in filters with longer effective wavelength, as expected due to evolutionary fading of the clusters strengthens the conclusion that the bend in the LF is connected to a maximum mass (by comparison with the models). The bright end slope of the LF is in all filters consistent with the slope expected from analytical arguments of \citet{gieles08}. 

The interpretation of the double power law distribution function is guided by the artificial population models and leads to the following conclusions:
\begin{enumerate} 
\item The bend in the LF occurs at \textit{brighter} magnitudes, \textit{closer} to the center of the galaxy. This may imply a higher possible maximum mass closer to the center of the galaxy. This conclusion is guided by the Monte Carlo simulations, which show that for given cluster formation history and disruption time, the bend magnitude is brighter for higher maximum masses. Varying the cluster formation history, and/or cluster disruption across the disk makes this inference less straightforward. However, assuming a constant maximum mass would require large variations in the disruption time as a function of galactocentric distance.
\item The magnitude of the bend in the LF varies slowly with background intensity. This variation may be solely due to the difference in average galactocentric distance between both background intensity regions.
\item The other deduction from the LF parameters concerns disruption parameters, which also depend on galactocentric distance. A higher surrounding density will be more destructive for a cluster, as encounters with massive objects, clouds and clusters, will be more frequent. Also, the tidal field of the galaxy is decreasing outwards. Due to the degeneracy of a decreasing upper mass limit and stronger disruption to flatten the faint end of the LF, the dependence of the disruption time on galactocentric distance can not be constrained. This degeneracy can be lifted if the ages and masses of the clusters can be measured.
\item The slope of the faint end side of the LF is \textit{shallower, closer} to the center of the galaxy and supports the suggestion that star clusters in the center are more easily disrupted. The higher density, stronger tidal field and larger relative velocities between several massive components (clusters and clouds) make the disruption of star clusters easier in those parts of the galaxy \citep{gieles06c, gieles06b}.
\item The slope of the faint end side of the LF is \textit{shallower} in \textit{high} background regions. This is counter-intuitive, as one expects the younger population in the spiral arms to have a faint-end LF slope more closely resembling the cluster IMF slope compared to the population in the interarm regions. Selection effects may play an important role here. Clusters have a high number density, causing confusion, blending and the rejection of clusters from the sample as a result of a very nearby neighbor. 
\end{enumerate}

Whether galactocentric distance or background intensity is the driving parameter for the variations of the faint-end slopes and bend magnitudes is unclear, due to the correlation of background intensity with galactocentric distance.

\begin{acknowledgements}
The anonymous referee is gratefully acknowledged for useful comments and suggestions. We kindly thank Marcelo Mora for providing us with the empirical PSF.
\end{acknowledgements}

\bibliographystyle{aa} 

\bibliography{20078831}

\appendix
\section{Comparison of fit methods} \label{sec:fits}

In order to fit a distribution function to the luminosity distribution of the clusters we exploit a method described and tested by \citet{maizapellaniz05}, who adopted it from \citet{dagostinostephens}. In conventional fit methods one uses equally sized bins (in either linear or logarithmic scale) for the quantity in question, luminosity in our case. Because the distribution is described by a power law with negative index, this always means that the bins at lower luminosity will contain more clusters and will therefore get a higher statistical weight in the fit (due to the Poissonian nature of the statistics in counting experiments). This results in an, on average, shallower slope from the fit routine than is actually present in the data, as shown by the Monte Carlo simulations of \citet{maizapellaniz05}. In order to overcome this problem these authors propose to use bins of variable binsize, chosen such that every bin contains equally many clusters and therefore have the same absolute Poissonian error. Besides the method of \citet{maizapellaniz05} we tested a cumulative distribution method, used by \citet{rosolowsky05} and a maximum likelihood method. 
\begin{figure*} 
   \centering
   \includegraphics[width=\textwidth]{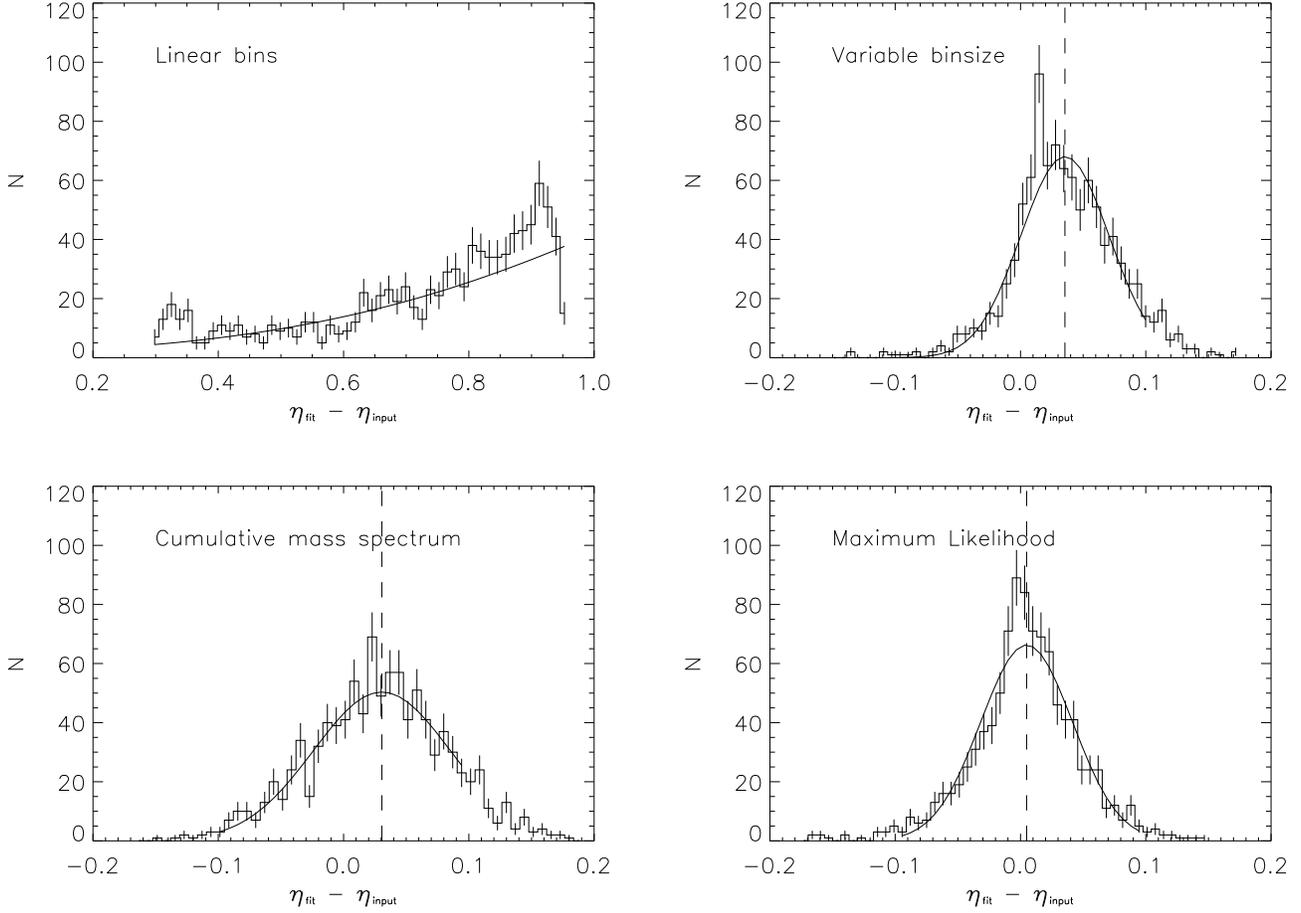}
      \caption{A comparison of four different methods to fit a distribution function to a data set. All frames are on the same scale, but the horizontal axis of the upper left panel is different. Histograms of the difference between the input slope and the fitted slope ($\eta_{\textrm{fit}} - \eta_{\textrm{input}}$, with $\eta_{\textrm{input}} = -2.35$) are shown for four different methods to fit a power law distribution function. Upper left: equally sized, linear bins. Upper right: bins with an equal number of objects per bin. Lower left: cumulative mass function. Lower right: Maximum Likelihood. In all panels the vertical dashed line indicates the center of the Gaussian that is fit to the distribution of fit results. See text for details on the methods.}
         \label{fig:fitcomp}
\end{figure*}

In order to make a comparison between the four methods to fit a distribution function, we follow a method similar to \citet{maizapellaniz05}.
For every method a sample of 10.000 numbers is randomly drawn from a power law distribution function with input slope -2.35, by linking a random number from a uniform distribution to the cumulative probability density function, belonging to the power law distribution function. This sample is fit by the four methods (described below) and the resulting slope is recorded (we fitted on the probability density function, such that there is only 1 free parameter: the slope). This process is repeated a 1000 times and the difference between the input slope ($\eta_{\rm input}$) and the slope that results from the fitting procedure ($\eta_{\rm fit}$) is plotted in Fig.~\ref{fig:fitcomp}.
 The methods: 
\begin{enumerate}
\item The `conventional method' uses equally sized bins (either linearly or logarithmically, linearly is shown in the upper left panel of Fig.~\ref{fig:fitcomp}). The problem of the difference in statistical weights makes the resulting histogram a function of the slope used for the distribution function (the steeper the function, the larger the difference in number of clusters in the low mass/luminosity bins and the greater the spread in Poissonian errors). In the case of a slope $\eta = -2.35$ this results in a large systematic offset as shown in Fig.~\ref{fig:fitcomp} (upper left panel).
\item \citet{maizapellaniz05} use a method with variable bin sizes, such that every bin contains equally many objects. The fit is consequently performed on the width of the bin, instead of the height. Errors in the widths are very small if photometric errors are neglected (ordered masses or luminosities are very close together for large samples), such that the Poissonian error on the number of clusters in the bin (the square root of that same number) is always dominant and the same for every bin. Using this kind of fitting results in residual slopes as shown in Fig.~\ref{fig:fitcomp} (upper right panel).
\item A method used, among others, by \citet{rosolowsky05}, involves no binning and makes use of the equation 
\begin{equation}
N(m \geq m') = C \int_{m'}^{\infty} N(m)\, \d m 
\end{equation}
(the cumulative distribution). The left hand side is the number of objects with a mass greater than or equal to $m'$ and $C$ is the normalization (from total number of clusters). While the fact of not having to deal with bins (and therefore Poissonian errors) is  ideal, the computational time necessary to deal with this method becomes long (quadratically with the number of objects), and therefore the lower left panel of Fig.~\ref{fig:fitcomp} is composed of samples with only 1000 objects, ten times less than the other ones. In this study, the number of clusters is large and therefore we will not use this method.
\item A reliable method that does not make use of bins is the maximum likelihood method \citep{bevington}. The likelihood that a certain guessed theoretical distribution function underlies the data is maximized. Here, the `guessed' distributions are power laws in which only the slope is changed in very small steps (0.005) from -0.35 to -4.35. 
The likelihood function $\mathcal{L}$ is given by
\begin{equation}
\mathcal{L} = \prod_{i} \, f(X_i)
\end{equation}
In here, $f$ is the function describing the guessed distribution, and $X_i$ are the values of the quantity one wants to fit (e.g. the luminosity), for every object $i$ in the sample. In the case of a power law distribution function for luminosities the likelihood function would be
\begin{equation}
\mathcal{L} = \prod_{i} \, L_i^{-\alpha}
\end{equation}
So for every input value of $\alpha$ the likelihood function will have a different value. The shape as a function of alpha will generally be gaussian. The peak of this gaussian defines the best fit for the power law slope, and the FWHM of the gaussian is the standard deviation. As the likelihood function generally consist of small numbers, it is common practice to maximize the logarithm of the likelihood function:
\begin{equation}
\log(\mathcal{L}) = -\alpha \cdot \sum_{i} \, \log(L_i)
\end{equation}
The maximum likelihood is found at offsets from the input value as shown in Fig.~\ref{fig:fitcomp} (lower right panel). Whereas this method in principle is the most reliable of the four, fitting several parameters simultaneously gives likelihood functions with several local minima, that may be comparable in significance. Whereas this would imply that a single good fit cannot be obtained, it is hard to automize the selection of the best fit. Therefore, and because the difference in fit quality with the method with variable bins is almost negligable, we choose not to use this method. 
\end{enumerate}

\noindent For consistency we always use the same method for fitting the distribution function. We choose for the second method, because we fit more than one parameters at once, in samples of thousands of clusters.

\end{document}